\documentclass[twocolumn,
superscriptaddress,
amssymb,
reprint,
nobibnotes,
aps,
prd]{revtex4-2}
\usepackage{hyperref}
\usepackage{amsmath}
\usepackage{enumitem}
\usepackage{graphicx}
\usepackage{subfigure}
\usepackage{tabularx}
\usepackage{multirow}
\usepackage{float}
\usepackage[dvipsnames,table]{xcolor}
\usepackage{color}
\hypersetup{
  colorlinks=true,
  urlcolor=NavyBlue,
  citecolor=Orchid,
  linkcolor=Magenta}

\newcommand{\keVnr}{$\text{keV}_\text{nr}$}
\newcommand{\eVnr}{$\text{eV}_\text{nr}$}
\newcommand{\eVee}{$\text{eV}_\text{ee}$}
\begin{document}

\title{Molecular Dynamics Simulations on Nuclear Recoils in Silicon
  Crystals towards Single Electron-Hole Pair Ionization Yields}

\author{Chang-Hao~Fang}
\affiliation{College of Physics, Sichuan University, Chengdu 610065}

\author{Shin-Ted~Lin}
\email[Corresponding Author: ]{stlin@scu.edu.cn}
\affiliation{College of Physics, Sichuan University, Chengdu 610065}

\author{Shu-Kui~Liu}
\email[Corresponding Author: ]{liusk@scu.edu.cn}
\affiliation{College of Physics, Sichuan University, Chengdu 610065}

\author{Henry~Tsz-King~Wong}
\affiliation{Institute of Physics, Academia Sinica, Taipei 11529}

\author{Hao-Yang~Xing}
\affiliation{College of Physics, Sichuan University, Chengdu 610065}

\author{Li-Tao~Yang}
\affiliation{Key Laboratory of Particle and Radiation Imaging
  (Ministry of Education) and Department of Engineering Physics,
  Tsinghua University, 100084 Beijing}

\author{Qian~Yue}
\affiliation{Key Laboratory of Particle and Radiation Imaging
  (Ministry of Education) and Department of Engineering Physics,
  Tsinghua University, 100084 Beijing}

\author{Jing-Jun~Zhu}
\affiliation{Institute of Nuclear Science and Technology, Sichuan
  University, Chengdu 610065}

\date{\today}%
\begin{abstract}
  We have developed a novel methodology utilizing molecular dynamics
  simulations to evaluate the ionization yields of nuclear recoils in
  crystalline silicon.
  This approach enables analytical exploration of atomic-scale transport
  within the lattice without necessitating parameterization.
  The quenching factors across the nuclear recoil energy range from 20 eV
  to 10 keV have been thoroughly investigated.
  A remarkable agreement with experimental data is achieved, particularly
  for the minimal energy regime conducted to date, reaching the level of a
  single electron-hole pair.
  This work presents a crucial and fundamental distribution of
  the quenching factor, which can be associated to the collisional
  interactions underlying the transport phenomena.
  The discrepancies observed with Lindhard's model for the quenching factor 
  at nuclear recoil energies below 4 keV are primarily attributed to lattice 
  binding effects and the specific characteristics of the crystal structure.
  In contrast, a gradual functional relationship is identified below
  approximately 100 eV, indicating that the quenching factor is influenced
  by the crystallographic orientation of the target material.
  From a distributional perspective, our analysis allows for the
  determination of the minimum exclusion mass for the dark matter nucleon
  elastic scattering channel at 0.29 $\mathrm{GeV}/c^2$, thereby
  significantly enhancing sensitivity for the sub-$\mathrm{GeV}/c^2$ mass
  region.
\end{abstract}
\maketitle

\emph{Introduction.}--
Coherent elastic scattering of neutrinos (CE$\nu$NS) and spin-independent 
Weak Interacting Massive Particles (WIMPs)~\cite{goodmanDetectabilityCertainDarkmatter1985} scattering off atomic nuclei 
mediated by the weak neutral current exhibit a notable advantage due to their substantial cross-sections compared to 
other weak interaction processes~\cite{freedmanCoherentEffectsWeak1974}.
This advantage arises from the constructive interference of the individual
nucleons within the target nucleus, leading to a fully quantum coherent
effect while transferring minuscule
momentum~\cite{kermanCoherencyNeutrinonucleusElastic2016}.
The ability to leverage the quantum coherent effect represents a valuable
asset in the study of neutrinos and the pursuit of dark matter, offering
enhanced experimental prospects and potential breakthroughs in our
understanding of the Standard Model and the new physics beyond the
Standard Model.
Nevertheless, an experimental challenge lies in the tiny momentum transfer
from neutrinos or WIMPs to the
nucleus~\cite{xuDetectionCalibrationLowEnergy2023}, resulting in the
observation of CE$\nu$NS recently~\cite{akimovObservationCoherentElastic2017,
  akimovMeasurementCoherentElastic2022}
and the searches of direct detection of light WIMPs~\cite{lewinReviewMathematicsNumerical1996,jiangLimitsLightWeakly2018,
  armengaudSearchesElectronInteractions2018,
  abdelhameedFirstResultsCRESSTIII2019,
  agneseSearchLowmassDark2019,
  aguilar-arevaloConfirmationSpectralExcess2024}
encountering significant uncertainties in the ionization and/or light
yields of nuclear recoils at low-energy detector response.

\begin{figure*}[tpb]
  \centering
  \includegraphics[width=0.85\linewidth]{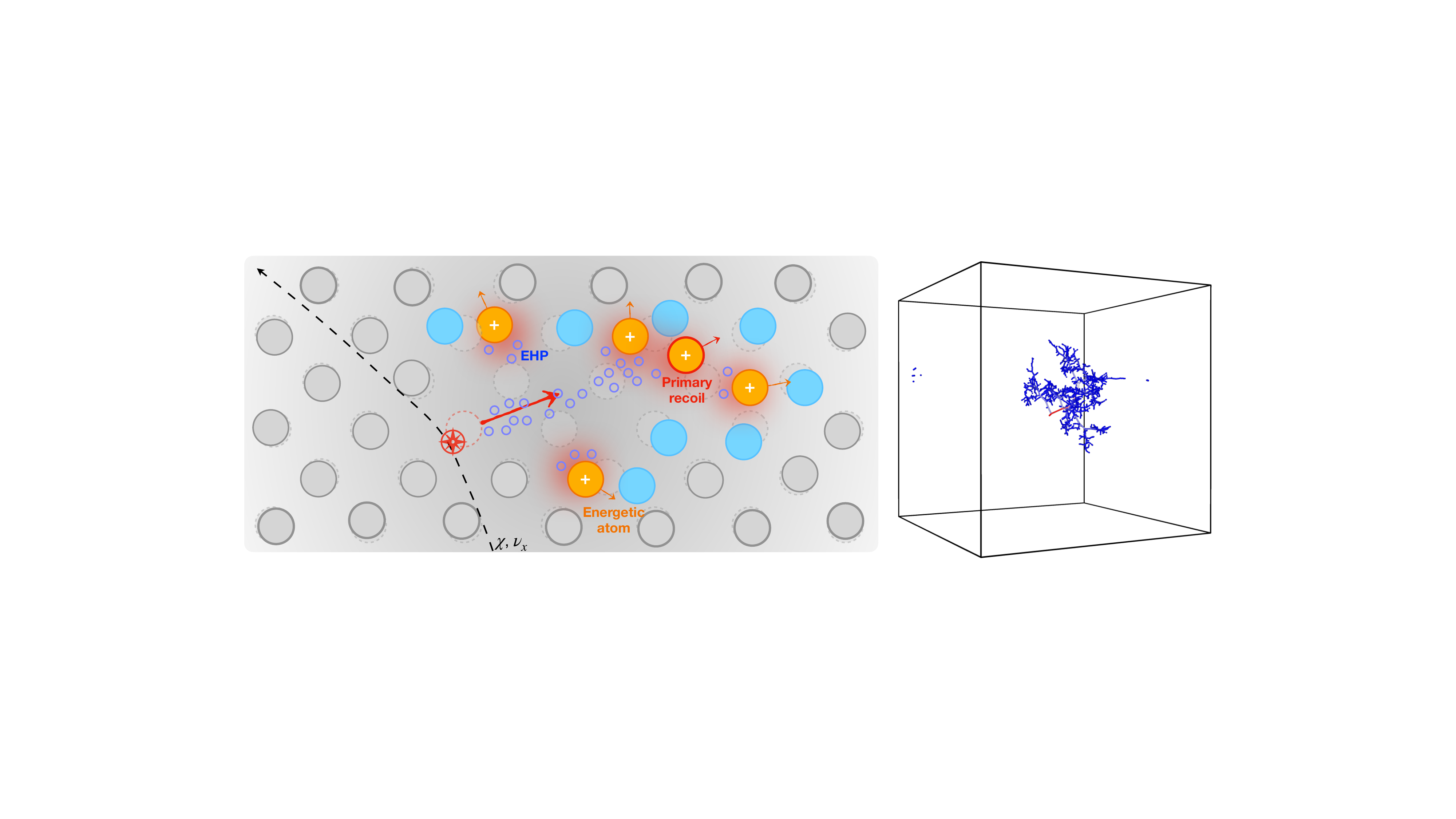}
  \caption{
    \textbf{[left]} The ionization production scenario involves the primary recoiled atom 
    (red edge), followed by displaced atoms (orange circles) that leave their lattice sites 
    during the cascade, and lattice-bound atoms (blue circles) that remain in position. 
    The grey fog and shadows represent electronic gas and the ionization process,
    respectively.
    The dashed circles indicate a perfect lattice without thermal relaxation. 
    \textbf{[right]} Traces of energetic atoms in a recoil event from MD 
    simulation.\label{fig:qf-scenario}}
\end{figure*}

The ``quenching factor'' (QF) is used to characterize and qualify the ratio
between the number of charge carriers generated by nuclear recoils and
electron recoils in a given material.
This work adopts ``$\mathrm{eV_{nr}}$''  and ``$\mathrm{eV_{ee}}$'' to represent 
nuclear recoil energy and electron equivalent ionization energy, respectively, 
unless explicitly noted otherwise.
The theoretical quenching model established by \citet{lindhard1963integral}
is considered the most successful, offering a concise formula for estimating the QF as 
a function of recoil energy.
The recent advancements in technology, such as the skipper charge-coupled device
(CCD)~\cite{tiffenbergSingleElectronSinglePhotonSensitivity2017,
  perezSearchingMillichargedParticles2024}, have enabled silicon to reach
the ionization energy threshold at the scale of a single
electron-hole pair (EHP).
In addition, the Super Cryogenic Dark Matter Search (SuperCDMS) Collaboration has advanced 
ionization yield measurements, achieving the lowest investigated energy of 
100 \eVnr~\cite{albakryFirstMeasurementNuclearRecoil2023}.
However, analysis of energy regimes below 10 keVnr indicates systematic overestimation of Lindhard's QF. 
This discrepancy becomes particularly pronounced below 4 \keVnr, where the QF overestimation exceeds 
20\% relative to experimental values. 
While \citet{sorensenAtomicLimitsSearch2015} and \citet{sarkisStudyIonizationEfficiency2020a}
attempted to address this through phenomenological parameterizations of atomic binding effects using existing 
QF measurements, their modified formulations fail to achieve predictive consistency, particularly in light of 
contradictory experimental data in the low-energy regime.

We have proposed an innovative approach utilizing molecular dynamics (MD)
simulations to characterize the QF over the low-energy range of 20 \eVnr{} to
10 \keVnr.
Unlike Lindhard's QF models, which yield a deterministic recoil energy function, our molecular dynamics 
approach accounts for probabilistic variations in monoenergetic recoil events.
This distribution arises from dynamic processes that 
incorporate the thermal lattice structure and advanced potentials tailored for silicon 
crystals.
By carefully considering specific factors relevant to low-energy scenarios,
our QF calculations align remarkably well with experimental data,
demonstrating accuracy even at the scale of a single EHP.

Accurate acknowledgment of the QFs is essential for advancing direct dark matter research.
The low-energy behavior of QF is crucial that involves ionization and scintillation techniques, 
directly influencing both detection efficiency and the physical interpretation of experimental results.
Single EHP technology in silicon targets enables exploration of sub-$\mathrm{GeV}/c^2$ 
WIMP masses.
Significantly, the state-of-the-art cryogenic calorimetry has achieved thresholds as low as a few 
\eVnr{}~\cite{changFirstLimitsLight2025, 
angloherFirstObservationSingle2024, 
goupyExploringCoherentElastic2023}, remaining unaffected by quenching factor effects.

\emph{Lindhard-like models.}--
\label{sec:pict-lindh-model}
Supposing the recoil energy $E_\mathrm{nr}$ is ultimately partitioned
between ionization production $\eta$ and atomic motion $\nu$,
\citet{lindhard1963integral} have proposed an approach to determine the
quenching factor.
The final average value of the atomic kinetic energy $\bar\nu$ is obtained by solving
the integral equation as follows:
\begin{equation}
  \label{eq:lindhard-homogen}
  k \varepsilon^{1 / 2} \bar{\nu}^{\prime}(\varepsilon)=\int_0^{\varepsilon^2} d t \frac{f\left(t^{1 / 2}\right)}{2 t^{3 / 2}} \times[\bar{\nu}(\varepsilon-t / \varepsilon)+\bar{\nu}(t / \varepsilon)-\bar{\nu}(\varepsilon)].
\end{equation}
Here, the parameter $k = 0.133 Z^{2/3} A^{-1/2}$ and the reduced recoil
energy $\varepsilon = c_Z E_\mathrm{nr}[\mathrm{keV}_\mathrm{nr}]$, 
where $c_Z = 11.5 / Z^{7/3}$.
Additionally, the parameter $t$ is proportional to the energy transfer to
atoms and is defined as $t\equiv\varepsilon^2\sin^2\theta$, where $\theta$ represents
the scattering angle in the center-of-mass frame.
Using the relation $\varepsilon = \bar\eta + \bar\nu$, the QF can be derived from
\begin{equation}
  \label{eq:lindhard-qf}
  \mathrm{QF}(\varepsilon) = \frac{\varepsilon - \bar\nu}{\varepsilon} = \frac{kg(\varepsilon)}{1+kg(\varepsilon)},
\end{equation}
where the semi-empirical function $g(\varepsilon)$ is often parameterized as
$g(\varepsilon) = 3\varepsilon^{0.15}+0.7\varepsilon^{0.6}+\varepsilon$~\cite{lewinReviewMathematicsNumerical1996}.


One approach to addressing the overestimation issues is to modify the 
assumptions in the original Lindhard model to suit low-energy scenarios. 
\citet{sorensenAtomicLimitsSearch2015} and
\citet{sarkisStudyIonizationEfficiency2020a,
  sarkisIonizationEfficiencyNuclear2023} reviewed these assumptions and 
highlighted that neglecting atomic binding effects (assumption (B) in 
Ref.~\cite{lindhard1963integral}) could be a 
significant reason for the overestimation.
Including the binding effect in their calculations allowed them to
achieve a lower QF than the original Lindhard's model.
However, the binding effect also introduced a pronounced ionization
threshold in the sub-\keVnr{} region, which is not observed in the current
measurement~\cite{albakryFirstMeasurementNuclearRecoil2023}.

\emph{Molecular Dynamics Approach.}--
\label{sec:princ-molec-dynam}
Classical MD calculates the time evolution of a system of
atoms by solving the Newtonian equations numerically.
This method is widely used for modeling recoil collision cascades in
materials, particularly at energies where multiple simultaneous
interactions play a significant
role.
It is considered an effective tool due to its good agreement with
experimental observations in various aspects,
such as ion range profile~\cite{nordlundLargeFractionCrystal2016,
  sillanpaaElectronicStoppingCalculated2001,
  sillanpaaElectronicStoppingSi2000a} and radiation defect
production~\cite{kitagawaIonirradiationExperimentFundamental1985,
  kirkCollapseDefectCascades1987,
  nordlundFormationStackingfaultTetrahedra1999,
  timonovaMolecularDynamicsSimulations2011}.

The ionization quenching process shares a physical picture
similar to the recoil collision cascades scenario.
Substantial advancements~\cite{nordAmorphizationMechanismDefect2002,
  timonovaMolecularDynamicsSimulations2011}, specifically in
semiconductors, spanning from \eVnr{} to \keVnr{} scales, indicate applicability
for our sub-\keVnr{} scale MD simulations.
In this work, we employ
\textsc{Lammps}~\cite{thompsonLAMMPSFlexibleSimulation2022,
  brownImplementingMolecularDynamics2013}, a widely used
classical MD simulation software, to model the ionization
quenching process of silicon.

The MD simulations calculated the dynamic processes as illustrated in
Fig.~\ref{fig:qf-scenario}.
The simulations commence with the ejection of an atom from the lattice
due to a recoil collision.
Subsequent interactions with surrounding atoms result in the deceleration
of the primary atoms (red circle) and the generation of collision cascades
(orange circles).
These cascades, along with the primary atom, contribute to the production
of ionizations (orange shadows).
The energy transferred to the electronic final state can subsequently be
segregated.

Interatomic potentials describe the interactions among atoms.
For covalent crystals like Si and Ge, the angular correlation
is necessary for potential to characterize their diamond cubic crystal 
structure~\cite{devanathanInteratomicPotentialsNuclear2020}.
The Tersoff potential is developed for covalent crystals and is widely
utilized in modeling radiation effects due to its reliability and
computational efficiency~\cite{tersoffNewEmpiricalApproach1988,
  nguyenGPUacceleratedTersoffPotentials2017}.
The Tersoff potential ensures adequately accurate interactions at
equilibrium distance, which produce the lattice
constant~\cite{tersoffNewEmpiricalApproach1988},
displacement threshold
energy~\cite{millerDisplacementthresholdEnergiesSi1994},
and cohesive energy~\cite{faridCohesiveEnergiesCrystals1991} with high
fidelity.
To address the severe overestimation for close encounter
interactions ($<1\ \text{\AA}$),
\citet{devanathanDisplacementThresholdEnergies1998} developed a Tersoff
potential for carbon and silicon, in which the short-range interactions are
used in conjunction with the Ziegler-Biersack-Littmark (ZBL) screened nuclear
repulsion potential~\cite{zieglerSRIMStoppingRange2015}.
Thanks to this, the behavior at equilibrium distance can be effectively
described using the Tersoff potential, while the hard collisions can
simultaneously be accurately captured using the ZBL potential.
In this work, we adopt this modified Tersoff potential as interatomic
interactions.

\begin{figure}[t]
  \centering
  \subfigure[] {
    \includegraphics[width=\linewidth]{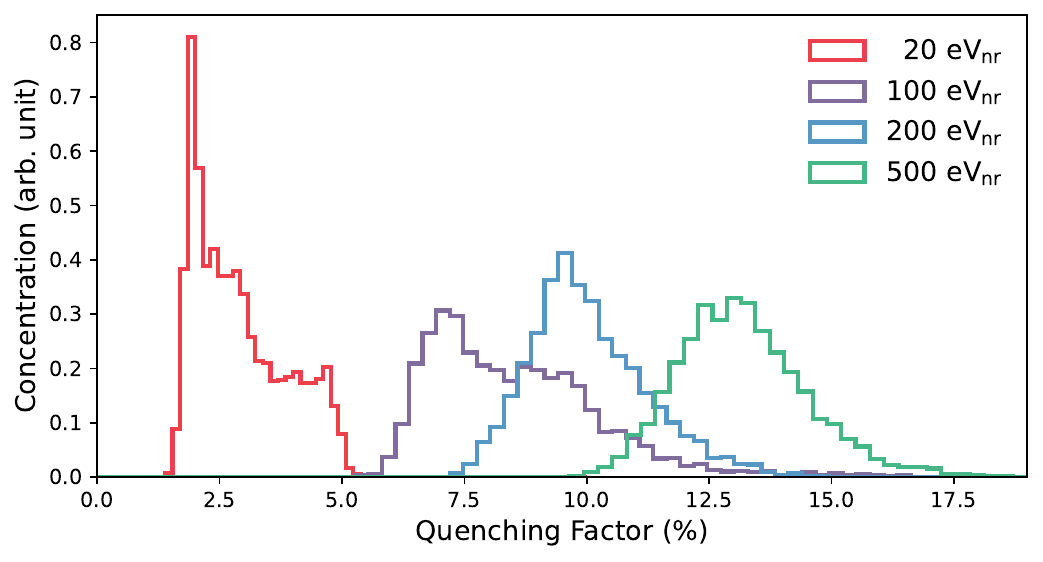}
    \label{fig:qf-dis-lowene}
  }
  \subfigure[] {
    \includegraphics[width=\linewidth]{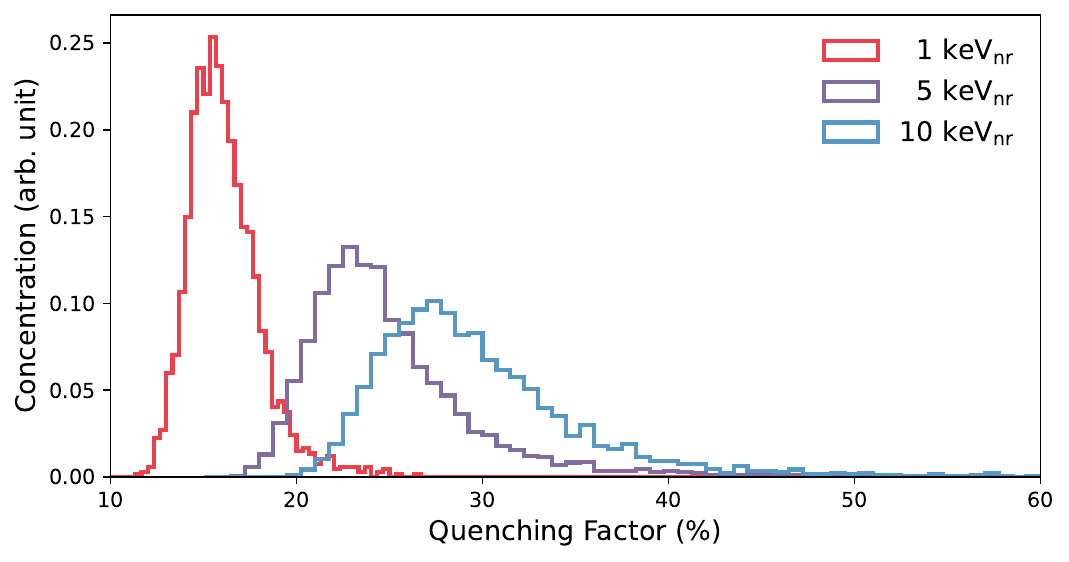}
    \label{fig:qf-dis-highene}
  }
  \caption{Inherent distributions of QFs.
    Figures~\ref{fig:qf-dis-lowene} and \ref{fig:qf-dis-highene} exhibit
    the dispersion properties in the sub-\keVnr{} and \keVnr{} regions,
    respectively.
    The expansion of the nonstructured QF distribution (above 200 \eVnr) is
    parameterized as $3.64 \times \mathrm{E[keV_{nr}]}^{0.43}$.}
  \label{fig:qf-dis}
\end{figure}

We incorporate atom-electron interactions based on the assumption that the nuclear and
electronic energy losses can be treated independently.
This approximation allows us to account for energy loss to electrons by
introducing the electronic stopping
power ($S_e$)~\cite{zieglerSRIMStoppingRange2015}, which has been validated by
range profile experiments~\cite{sillanpaaElectronicStoppingCalculated2001}.
\textsc{Lammps} introduces the inelastic electronic energy loss as a
friction force that decelerates energetic atoms based on the
$S_e$.
For each atom that suffers electronic stopping, an additional force is
applied as
\begin{equation}
  \label{eq:lmp-estop}
  \mathbf{F}_i = \mathbf{F}^0_i - \frac{\mathbf{v}_i}{\|\mathbf{v}_i\|} \cdot S_e(E),
\end{equation}
where $\mathbf{F}_i$, $\mathbf{F}^0_i$, and $\mathbf{v}_i$ represent the
total force, original force, and velocity of the $i^{th}$ atom,
respectively.
The $S_e$ is determined through a semi-empirical model
in Stopping and Range of Ions in Matter (\textsc{Srim}) 
software~\cite{zieglerSRIMStoppingRange2015}, 
as input for \textsc{Lammps}.
For silicon, the \textsc{Srim} $S_e$ is extensively constrained through direct
measurements spanning over decades.
Particularly, recent measurements robustly support the
low-energy extrapolation of the
model~\cite{lohmannTrajectorydependentElectronicExcitations2020}.

A simulation box with dimensions of $50\times50\times50$ ($20\times20\times20$) lattice units
is constructed, providing sufficient undisturbed lattice to accommodate
ion transport with kinetic energies below 10 \keVnr{}(6 \keVnr) under periodic
boundary conditions.
To achieve thermal equilibrium prior to recoil, thermal
relaxation is performed under the canonical ensemble at 300 K for
30 ps, utilizing a time step of 1 fs.

The recoiled atom, randomly selected from the thermalized crystal, is
ejected isotropically with a given recoil energy.
Then, the intermediate dynamic process is evaluated through MD simulations.
The simulation is halted when the energy of the most energetic atom
falls below 10 \eVnr, as no ionization occurs below this
threshold~\cite{limElectronElevatorExcitations2016}.
Ionization of this event is obtained by integrating $S_e$ along the
trajectories of all cascades.
The QF can then be derived from
\begin{equation}
  \label{eq:md-qf}
    I = \sum_{i=0}^{n_c}\int_{{\mathbf{r}_0}_i}^{{\mathbf{r}_1}_i}S_e[E_i(r_i)]\cdot dr_i, \quad
    \mathrm{QF} = \frac{I}{E_{\mathrm{nr}}},
\end{equation}
where ${{\mathbf{r}_0}_i}, {{\mathbf{r}_1}_i}$ represent an atom's initial
and final position, and $n_c$ denotes the number of energetic atoms.

\label{sec:results-discussion}
Intensive investigations of the QF have been conducted for various recoiled
energies.
Special attention is given to the low-energy region, particularly below 4
\keVnr, where Lindhard's model provides overestimated
predictions~\cite{izraelevitchMeasurementIonizationEfficiency2017a, 
albakryFirstMeasurementNuclearRecoil2023}.

\begin{figure*}[tp]
  \centering
  \includegraphics[width=0.8\textwidth]{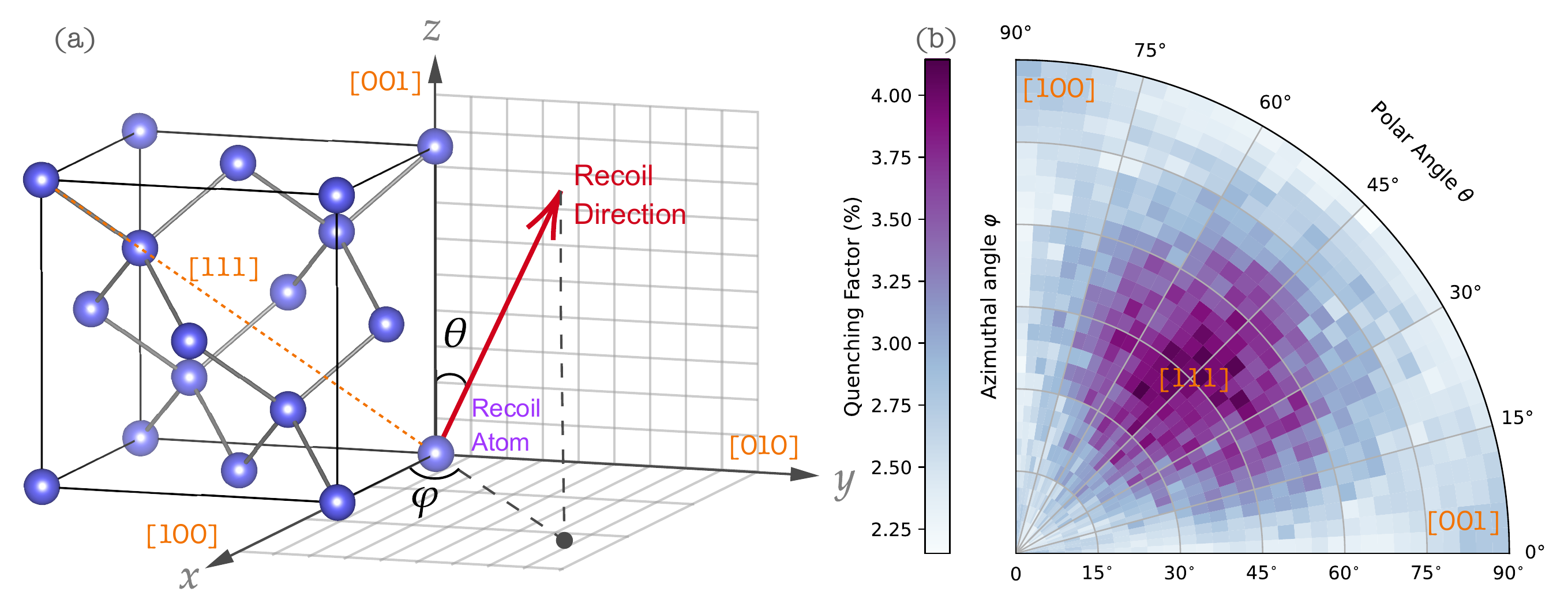}
  \caption{\textbf{[left]} Schematic of Si lattice with crystallographic directions 
    ([100], [001], [111]) and recoil angles $\theta/\varphi$ labeled.
    \textbf{[right]}
    Recoil angular dependence of the QF at 20 \eVnr, showing mean QF over $(\theta,\varphi)$.
    Attributed to the symmetry of Si, the $\theta$ and $\varphi$ can be
    reduced to $(0, 90^\circ)$.\label{fig:qf-angle-dis}}
\end{figure*}

\emph{Inherently Distributions.}--\label{sec:inher-fluct-qf}
As illustrated in Fig.~\ref{fig:qf-dis}, the complex process of multiple
collisions involved in slowing down an isotropically ejected atom does not
yield a single, definitive value; instead, it results in a distribution.
It is crucial to emphasize that the expansion of these distributions is not
sufficiently small to be ignored.
It necessitates further consideration when interpreting experimental
results, particularly in the context of rare events experiments.

The expansion of the QF tends to increase with recoil energies.
We employ the 1~$\sigma$ central confidence interval to describe 
the expansion of the QF distributions quantitatively.
For example, at 200 \eVnr, the 1~$\sigma$ region of QF spans from 8.88\% 
to 11.24\%, while at 10 \keVnr, it extends from 24.93\% to 34.33\%.
An empirical formula is used to approximate the energy dependence of the
1 $\sigma$ width of QF, expressed as ${3.64 \times E[\mathrm{keV_{nr}}]}^{0.43}$.
We note that the shapes of these distributions are not
identical.
Below 200 \eVnr, a platform-like structure emerges in the tail of the QF
distributions.
However, for energies above 200 \eVnr, the spectra exhibit a smoother shape.
In the following discussions, these two phenomena will be analyzed
separately.

When the recoil energy falls below 200 \eVnr, the atom-atom interactions are
no longer solely dominated by the ZBL repulsive potential.
Instead, a significant contribution from solid-state interactions,
specifically the Tersoff potential, becomes relevant.
The lattice binding, involved by the Tersoff potential, is approximately
10--20 eV~\cite{millerDisplacementthresholdEnergiesSi1994,
  holmstromThresholdDefectProduction2008} and varies depending on the
crystallographic orientations.
The highly anisotropic lattice binding, in conjunction with the crystal
structure, results in a pronounced angular dependence on QF, as depicted in
Fig.~\ref{fig:qf-angle-dis}(b) for a recoil energy of 20 \eVnr.
This reveals an ionization enhancement in specific directions, leading
to the platform-like structures depicted in Fig.~\ref{fig:qf-dis-lowene}.
Furthermore, our approach naturally incorporates directional
lattice binding rather than assuming it as a definitive parameter.
Consequently, our calculations do not impose a strong threshold on QF
predictions, as seen in the Lindhard-like models.
The notable contribution from the crystal effect addresses a limitation of
the Lindhard-like model, which completely ignores interactions at the
energy scale of solid-state physics.
By incorporating equilibrium interactions and crystal structure, we
observe distinct QF behavior in the extremely low-energy region.

For recoil energies above 200 \eVnr, the previously mentioned influences
gradually diminish with increasing recoil energies.
As illustrated in Fig.~\ref{fig:qf-dis-highene}, the distinct structures
and angular dependence observed at energies below 200 \eVnr{} are no longer
present.
These can be attributed to the relatively small energy scale of crystal
binding in comparison to the current recoil energy.
Additionally, the relatively long and random trajectories of the recoiling
atoms tend to average out the effects of the crystal structure.

\emph{Experimental investigation.}--
Investigating QF behavior across thermal environments from 52 millikelvin (mK) to 300 K, 
we find a minor enhancement with increasing temperature. 
The temperature-induced variations are more pronounced at lower energies. 
Quantitatively, at 20 \eVnr, the QF at 52 mK is approximately 92.7\% of its value at 300 K. 
However, for recoil energies above 1 \keVnr, the relative difference between 52 mK and 300 K 
diminishes to less than 0.7\%.
These findings align with our expectations, as the impact of temperature on
atomic kinetic energy ($10^{-2}$ \eVnr{} at 300 K, $10^{-6}$ \eVnr{} at
52 mK) is several orders of magnitude lower than the recoil energy.

\begin{figure}[tp]
  \centering
  \includegraphics[width=\linewidth]{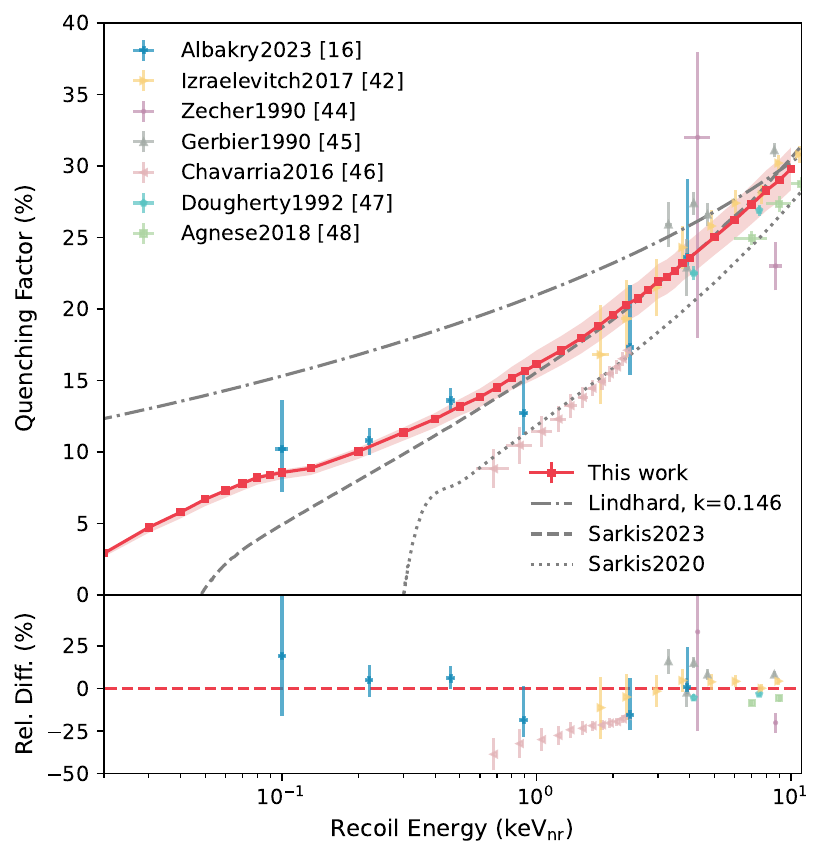}
  \caption{\textbf{[top]} MD-simulated silicon quenching factors
    (red points; the systematic uncertainties are present with a red shaded area) are
    compared with experimental measurements (colored crosses)~\cite{izraelevitchMeasurementIonizationEfficiency2017a,
      zecherEnergyDepositionEnergetic1990,
      albakryFirstMeasurementNuclearRecoil2023,
      gerbierMeasurementIonizationSlow1990,
      chavarriaMeasurementIonizationProduced2016,
      doughertyMeasurementsIonizationProduced1992,
      agneseNuclearrecoilEnergyScale2018} with Lindhard-like models
    superimposed (gray dashed lines, including the original Lindhard model
    from Eq.~\eqref{eq:lindhard-qf})~\cite{lindhard1963integral,
      sarkisIonizationEfficiencyNuclear2023,
      sarkisStudyIonizationEfficiency2020a}.
    \textbf{[bottom]} The relative difference between our calculations and
    measurements.}
  \label{fig:qf-exp-model}
\end{figure}

\label{sec:molec-dynam-quench}
In Fig.~\ref{fig:qf-exp-model}, we compare our QF, the experimentally measured QF
from
references~\cite{izraelevitchMeasurementIonizationEfficiency2017a,
      zecherEnergyDepositionEnergetic1990,
      albakryFirstMeasurementNuclearRecoil2023,
      gerbierMeasurementIonizationSlow1990,
      chavarriaMeasurementIonizationProduced2016,
      doughertyMeasurementsIonizationProduced1992,
      agneseNuclearrecoilEnergyScale2018}, and the predictions of
Lindhard-like models~\cite{lindhard1963integral,
  sarkisStudyIonizationEfficiency2020a,
  sarkisIonizationEfficiencyNuclear2023}.
Each data point (red points) represents the mean value obtained from the
corresponding distributions discussed earlier.
The simulations are conducted on a sufficiently large scale to minimize the statistical error, 
which is estimated to be within 0.13\%.
Systematic error sources include interatomic potential, $S_e$, temperature, and the simulation 
parameterization framework, with Se contributing most significantly. 
The lowest-energy $S_e$ measurement introduces $\sim${}7\% uncertainty, propagating to $\le$ 6\% relative uncertainty. 
Temperature deviations from experimental conditions cause $\le$ 4\% differences above 100 \eVnr, while system size effects 
are minimal ($<0.53\%$). 
The potential energy function's impact remains unquantified due to complexity, though its widespread 
use suggests a limited effect. 
Total uncertainties, assuming propagated independently, are shown as the shaded region in Fig.~\ref{fig:qf-exp-model}.

Our calculations demonstrate excellent agreement with experimental
measurements below 10 \keVnr, surpassing the performance of other models,
particularly for energies below 4 \keVnr.
The results align well with recent measurements conducted by the SuperCDMS
Collaboration ($\chi^2/\mathrm{n.d.f.} = 8.61/6$), which reported a lowest recoil energy of 
100 \eVnr{} using a monochromatic neutron
facility~\cite{albakryFirstMeasurementNuclearRecoil2023}.
Notably, our calculations demonstrate the best agreement with current observations.
This achievement supports the reliability of our MD calculations even to
the level of the single EHP creation.
While our average QF results match monochromatic recoil data well, meaningful comparison 
with continuous-spectrum measurements, discrepancies arise when comparing them 
to continuous-spectrum measurements, particularly with the experimental data 
from \citet{chavarriaMeasurementIonizationProduced2016}.
A meaningful comparison necessitates the convolution of the neutron spectrum and the detector 
response, but that lies beyond the scope of this study.

Two critical features are obtained in our calculations.
Firstly, our calculations yield QFs that are lower than the prediction from
Lindhard's model.
This can be attributed to the lattice binding effects, in tune with the
discussions of Lindhard-like models~\cite{sorensenAtomicLimitsSearch2015,
  sarkisStudyIonizationEfficiency2020a}.
Secondly, as illustrated in Fig.~\ref{fig:qf-exp-model}, a non-trivial
transition occurs around 100 \eVnr{} and below.
The transition is a result of the growing impact of the crystal structure
and non-homogeneity of lattice binding, which leads to increased ionization
in certain directions and causes non-trivial behavior around 100 \eVnr.

The fluctuation in ionization measurements for near single EHP is
determined using a variance-modified Poisson statistical
model~\cite{tiffenbergSingleElectronSinglePhotonSensitivity2017,
  perezSearchingMillichargedParticles2024,
  romaniThermalDetectionSingle2018}, which is represented by the Fano
factor~\cite{fanoIonizationYieldRadiations1947}.
Recent analyses~\cite{mathenyIntrinsicFanoFactor2022,
  albakryFirstMeasurementNuclearRecoil2023} reveal significant smearing
in the Fano factor to ionization yields~\cite{mazziottaElectronHolePair2008} and optical photon
generation~\cite{bousselhamPhotoelectronAnticorrelationsSubPoisson2010}
from nuclear recoils, suggesting a distribution of QF.
Further quantitative analysis remains necessary.

We evaluate the influence of QF models on the interpretation of dark matter
search results.
The differential event rate $\mathrm{d}R/\mathrm{d}E_\mathrm{nr}$ for
spin-independent dark
matter-nucleon ($\chi$-N) couplings is derived using standard galactic halo
parameters in the elastic scattering model, with DM density
$\rho_\chi=0.3\ \mathrm{GeV/c^2/cm^3}$, escape velocity
$v_\mathrm{esc}=544\ \mathrm{km/s}$, and most probable velocity
$v_0 = 220\ \mathrm{km/s}$~\cite{lewinReviewMathematicsNumerical1996,
  baxterRecommendedConventionsReporting2021}.
For ionization detection, the measurable electronic equivalent energy
$E_\mathrm{ee} = \mathrm{QF}(E_\mathrm{nr}) \times E_\mathrm{nr}$.

\begin{figure}[t]
  \centering
  \includegraphics[width=\linewidth]{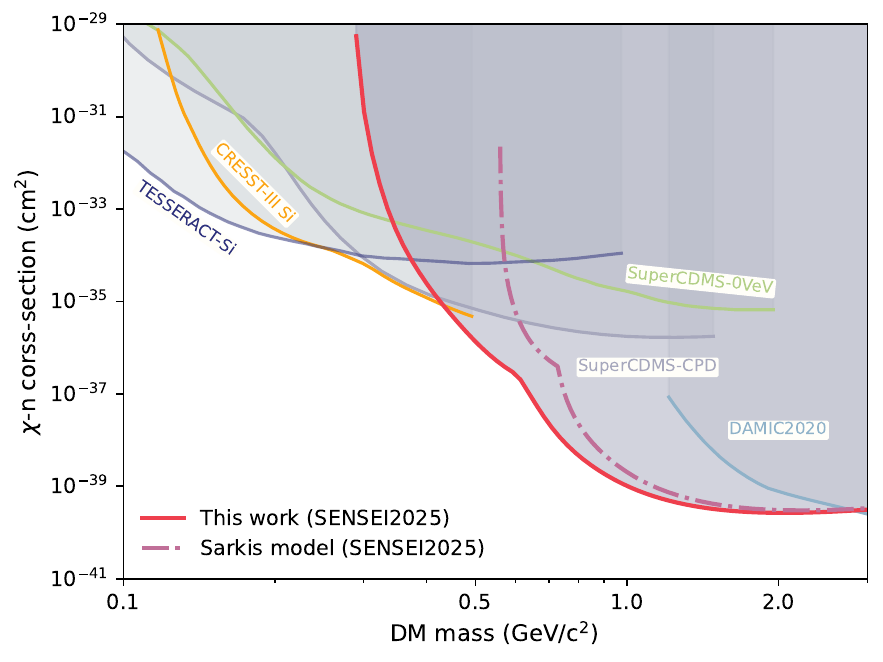}
  \caption{90\% C.L. upper limits for the $\chi$-N interaction derived from
    SENSEI
    spectrum~\cite{adariFirstDirectDetectionResults2025} using our
    distribution perspective QFs and the Sarkis QF
    model~\cite{sarkisIonizationEfficiencyNuclear2023}, along with results
    from silicon-based
    experiment~\cite{albakryInvestigatingSourcesLowenergy2022,
      alkhatibLightDarkMatter2021,
      angloherResultsSubGeVDark2023,
      aguilar-arevaloResultsLowMassWeakly2020,
      changFirstLimitsLight2025}.\label{fig:DM-limist}}
\end{figure}

Two approaches are employed to incorporate the ionization quenching effect.
The first treats the QF as a single value parameterized by recoil energy,
which can be achieved through variable substitution.
The second accounts for the irreducible energy-dependent intrinsic
distribution of the QF, in which case, the event rate becomes the
convolution of the QF distribution with the differential nuclear recoil
event rate.
Both QF treatments were employed to estimate exclusion limits using the
binned Poisson method~\cite{savageCompatibilityDAMALIBRA2009}, applied to
the most recent SENSEI
spectrum~\cite{adariFirstDirectDetectionResults2025}.
Fig.~\ref{fig:DM-limist} illustrates the 90\% confidence level (C.L.)
upper limit for spin-independent $\chi$-N couplings in
the silicon targets~\cite{aguilar-arevaloConfirmationSpectralExcess2024,
      albakryInvestigatingSourcesLowenergy2022,
      alkhatibLightDarkMatter2021,
      angloherResultsSubGeVDark2023,
      aguilar-arevaloResultsLowMassWeakly2020,
      changFirstLimitsLight2025}.
Our QF interpretation, characterized by a larger mean value and an
asymmetric tail distribution, evidently enhances the exclusion
limits at sub-$\mathrm{GeV}/\mathrm{c}^2$ region and extends the excluded
ability of the  $\chi$-N channel to a mass of 0.29 $\mathrm{GeV}/\mathrm{c}^2$, corresponding to 
a 6.2 \eVee{} analysis
threshold~\cite{adariFirstDirectDetectionResults2025}.
These effects are particularly pronounced in the search for
sub-$\mathrm{GeV}/\mathrm{c}^2$ WIMP masses.

\emph{Conclusions.}--\label{sec:conclusions}
The MD simulation, which incorporates many-body effects, provides a
comprehensive understanding of the explicit interactions among atoms in
crystalline silicon.
Notably, the lattice binding energy and multiple collision effects are
intrinsically associated with the atomic-scale transport.
This analysis leads to the emergence of a novel perspective, wherein the QF
is regarded as a distribution, rather than a singular value,
generated through the collision process.
Furthermore, it demonstrates an excellent agreement with low-energy nuclear
recoil measurements in silicon, especially for the minimal recoil energy of
100 \eVnr{} as conducted by the SuperCDMS 
Collaboration~\cite{albakryFirstMeasurementNuclearRecoil2023}.

In contrast to the Lindhard-like models, the MD simulation approach offers an
intuitive evaluation of the QF by integrating the silicon lattice binding
energy with atomic collision processes.
The electronic final state model, grounded in extensive experimental
data~\cite{zieglerSRIMStoppingRange2015,
  lohmannTrajectorydependentElectronicExcitations2020} and first-principle
calculations~\cite{limElectronElevatorExcitations2016},
highlights the presence of ultra-low ionization thresholds resulting from
the effects of dynamic defects.
The Lindhard model is shown to overestimate QF for recoil energies below
approximately 4 \keVnr, primarily due to the impact of lattice binding
energy.
Furthermore, the directionality of the crystal structure dependence
leads to significant sensitivity at energies below 200 \eVnr.
It is these features that demonstrate unanimous agreement with the data
obtained from the EHP counting detector.
Additionally, no significant temperature dependence was observed across all
energy levels.

Understanding the detector response to low-energy recoils is crucial for
investigating low-mass WIMPs and CE$\nu$NS.
We utilize the SENSEI data~\cite{adariFirstDirectDetectionResults2025}
with the new QF results to place constraints on the $\chi$-N
interaction, which improves upon the previous bounds in the mass region
of 0.8--2 $\mathrm{GeV/c}^2$ and extends the minimum exclusion mass to 0.29
$\mathrm{GeV/c}^2$ in silicon-based experiments.
The MD method represents a unique and vital approach for estimating the
transport processes of recoil nuclei, with potential applications in
other detection techniques, such as light yield measurements.

\begin{acknowledgments}
  This work was supported by the National Key Research and Development
  Program of China (Contract No. 2023YFA1607103) and the National Natural
  Science Foundation of China (Contracts No. 12441512, No. 11975159, No.
  11975162) provided support for this work.
\end{acknowledgments}

\bibliography{ref.bib}

\begin{thebibliography}{60}%
\makeatletter
\providecommand \@ifxundefined [1]{%
 \@ifx{#1\undefined}
}%
\providecommand \@ifnum [1]{%
 \ifnum #1\expandafter \@firstoftwo
 \else \expandafter \@secondoftwo
 \fi
}%
\providecommand \@ifx [1]{%
 \ifx #1\expandafter \@firstoftwo
 \else \expandafter \@secondoftwo
 \fi
}%
\providecommand \natexlab [1]{#1}%
\providecommand \enquote  [1]{``#1''}%
\providecommand \bibnamefont  [1]{#1}%
\providecommand \bibfnamefont [1]{#1}%
\providecommand \citenamefont [1]{#1}%
\providecommand \href@noop [0]{\@secondoftwo}%
\providecommand \href [0]{\begingroup \@sanitize@url \@href}%
\providecommand \@href[1]{\@@startlink{#1}\@@href}%
\providecommand \@@href[1]{\endgroup#1\@@endlink}%
\providecommand \@sanitize@url [0]{\catcode `\\12\catcode `\$12\catcode `\&12\catcode `\#12\catcode `\^12\catcode `\_12\catcode `\%12\relax}%
\providecommand \@@startlink[1]{}%
\providecommand \@@endlink[0]{}%
\providecommand \url  [0]{\begingroup\@sanitize@url \@url }%
\providecommand \@url [1]{\endgroup\@href {#1}{\urlprefix }}%
\providecommand \urlprefix  [0]{URL }%
\providecommand \Eprint [0]{\href }%
\providecommand \doibase [0]{https://doi.org/}%
\providecommand \selectlanguage [0]{\@gobble}%
\providecommand \bibinfo  [0]{\@secondoftwo}%
\providecommand \bibfield  [0]{\@secondoftwo}%
\providecommand \translation [1]{[#1]}%
\providecommand \BibitemOpen [0]{}%
\providecommand \bibitemStop [0]{}%
\providecommand \bibitemNoStop [0]{.\EOS\space}%
\providecommand \EOS [0]{\spacefactor3000\relax}%
\providecommand \BibitemShut  [1]{\csname bibitem#1\endcsname}%
\let\auto@bib@innerbib\@empty
\bibitem [{\citenamefont {Goodman}\ and\ \citenamefont {Witten}(1985)}]{goodmanDetectabilityCertainDarkmatter1985}%
  \BibitemOpen
  \bibfield  {author} {\bibinfo {author} {\bibfnamefont {M.~W.}\ \bibnamefont {Goodman}}\ and\ \bibinfo {author} {\bibfnamefont {E.}~\bibnamefont {Witten}},\ }\href {https://doi.org/10.1103/PhysRevD.31.3059} {\bibfield  {journal} {\bibinfo  {journal} {Phys. Rev. D}\ }\textbf {\bibinfo {volume} {31}},\ \bibinfo {pages} {3059} (\bibinfo {year} {1985})}\BibitemShut {NoStop}%
\bibitem [{\citenamefont {Freedman}(1974)}]{freedmanCoherentEffectsWeak1974}%
  \BibitemOpen
  \bibfield  {author} {\bibinfo {author} {\bibfnamefont {D.~Z.}\ \bibnamefont {Freedman}},\ }\href {https://doi.org/10.1103/PhysRevD.9.1389} {\bibfield  {journal} {\bibinfo  {journal} {Phys. Rev. D}\ }\textbf {\bibinfo {volume} {9}},\ \bibinfo {pages} {1389} (\bibinfo {year} {1974})}\BibitemShut {NoStop}%
\bibitem [{\citenamefont {Kerman}\ \emph {et~al.}(2016)\citenamefont {Kerman} \emph {et~al.}}]{kermanCoherencyNeutrinonucleusElastic2016}%
  \BibitemOpen
  \bibfield  {author} {\bibinfo {author} {\bibfnamefont {S.}~\bibnamefont {Kerman}} \emph {et~al.},\ }\href {https://doi.org/10.1103/PhysRevD.93.113006} {\bibfield  {journal} {\bibinfo  {journal} {Phys. Rev. D}\ }\textbf {\bibinfo {volume} {93}},\ \bibinfo {pages} {113006} (\bibinfo {year} {2016})}\BibitemShut {NoStop}%
\bibitem [{\citenamefont {Xu}\ \emph {et~al.}(2023)\citenamefont {Xu} \emph {et~al.}}]{xuDetectionCalibrationLowEnergy2023}%
  \BibitemOpen
  \bibfield  {author} {\bibinfo {author} {\bibfnamefont {J.}~\bibnamefont {Xu}} \emph {et~al.},\ }\href {https://doi.org/10.1146/annurev-nucl-111722-025122} {\bibfield  {journal} {\bibinfo  {journal} {Annu. Rev. Nucl. Part. Sci.}\ }\textbf {\bibinfo {volume} {73}},\ \bibinfo {pages} {95} (\bibinfo {year} {2023})}\BibitemShut {NoStop}%
\bibitem [{\citenamefont {Akimov}\ \emph {et~al.}(2017)\citenamefont {Akimov} \emph {et~al.}}]{akimovObservationCoherentElastic2017}%
  \BibitemOpen
  \bibfield  {author} {\bibinfo {author} {\bibfnamefont {D.}~\bibnamefont {Akimov}} \emph {et~al.} (\bibinfo {collaboration} {COHERENT Collaboration}),\ }\href {https://doi.org/10.1126/science.aao0990} {\bibfield  {journal} {\bibinfo  {journal} {Science}\ }\textbf {\bibinfo {volume} {357}},\ \bibinfo {pages} {1123} (\bibinfo {year} {2017})}\BibitemShut {NoStop}%
\bibitem [{\citenamefont {Akimov}\ \emph {et~al.}(2022)\citenamefont {Akimov} \emph {et~al.}}]{akimovMeasurementCoherentElastic2022}%
  \BibitemOpen
  \bibfield  {author} {\bibinfo {author} {\bibfnamefont {D.}~\bibnamefont {Akimov}} \emph {et~al.} (\bibinfo {collaboration} {COHERENT Collaboration}),\ }\href {https://doi.org/10.1103/PhysRevLett.129.081801} {\bibfield  {journal} {\bibinfo  {journal} {Phys. Rev. Lett.}\ }\textbf {\bibinfo {volume} {129}},\ \bibinfo {pages} {081801} (\bibinfo {year} {2022})}\BibitemShut {NoStop}%
\bibitem [{\citenamefont {Lewin}\ \emph {et~al.}(1996)\citenamefont {Lewin} \emph {et~al.}}]{lewinReviewMathematicsNumerical1996}%
  \BibitemOpen
  \bibfield  {author} {\bibinfo {author} {\bibfnamefont {J.}~\bibnamefont {Lewin}} \emph {et~al.},\ }\href {https://doi.org/10.1016/S0927-6505(96)00047-3} {\bibfield  {journal} {\bibinfo  {journal} {Astroparticle Physics}\ }\textbf {\bibinfo {volume} {6}},\ \bibinfo {pages} {87} (\bibinfo {year} {1996})}\BibitemShut {NoStop}%
\bibitem [{\citenamefont {Jiang}\ \emph {et~al.}(2018)\citenamefont {Jiang} \emph {et~al.}}]{jiangLimitsLightWeakly2018}%
  \BibitemOpen
  \bibfield  {author} {\bibinfo {author} {\bibfnamefont {H.}~\bibnamefont {Jiang}} \emph {et~al.} (\bibinfo {collaboration} {CDEX Collaboration}),\ }\href {https://doi.org/10.1103/PhysRevLett.120.241301} {\bibfield  {journal} {\bibinfo  {journal} {Phys. Rev. Lett.}\ }\textbf {\bibinfo {volume} {120}},\ \bibinfo {pages} {241301} (\bibinfo {year} {2018})}\BibitemShut {NoStop}%
\bibitem [{\citenamefont {Armengaud}\ \emph {et~al.}(2018)\citenamefont {Armengaud} \emph {et~al.}}]{armengaudSearchesElectronInteractions2018}%
  \BibitemOpen
  \bibfield  {author} {\bibinfo {author} {\bibfnamefont {E.}~\bibnamefont {Armengaud}} \emph {et~al.} (\bibinfo {collaboration} {EDELWEISS Collaboration}),\ }\href {https://doi.org/10.1103/PhysRevD.98.082004} {\bibfield  {journal} {\bibinfo  {journal} {Phys. Rev. D}\ }\textbf {\bibinfo {volume} {98}},\ \bibinfo {pages} {082004} (\bibinfo {year} {2018})}\BibitemShut {NoStop}%
\bibitem [{\citenamefont {Abdelhameed}\ \emph {et~al.}(2019)\citenamefont {Abdelhameed} \emph {et~al.}}]{abdelhameedFirstResultsCRESSTIII2019}%
  \BibitemOpen
  \bibfield  {author} {\bibinfo {author} {\bibfnamefont {A.~H.}\ \bibnamefont {Abdelhameed}} \emph {et~al.} (\bibinfo {collaboration} {CRESST Collaboration}),\ }\href {https://doi.org/10.1103/PhysRevD.100.102002} {\bibfield  {journal} {\bibinfo  {journal} {Phys. Rev. D}\ }\textbf {\bibinfo {volume} {100}},\ \bibinfo {pages} {102002} (\bibinfo {year} {2019})}\BibitemShut {NoStop}%
\bibitem [{\citenamefont {Agnese}\ \emph {et~al.}(2019)\citenamefont {Agnese} \emph {et~al.}}]{agneseSearchLowmassDark2019}%
  \BibitemOpen
  \bibfield  {author} {\bibinfo {author} {\bibfnamefont {R.}~\bibnamefont {Agnese}} \emph {et~al.} (\bibinfo {collaboration} {SuperCDMS Collaboration}),\ }\href {https://doi.org/10.1103/PhysRevD.99.062001} {\bibfield  {journal} {\bibinfo  {journal} {Phys. Rev. D}\ }\textbf {\bibinfo {volume} {99}},\ \bibinfo {pages} {062001} (\bibinfo {year} {2019})}\BibitemShut {NoStop}%
\bibitem [{\citenamefont {{Aguilar-Arevalo}}\ \emph {et~al.}(2024)\citenamefont {{Aguilar-Arevalo}} \emph {et~al.}}]{aguilar-arevaloConfirmationSpectralExcess2024}%
  \BibitemOpen
  \bibfield  {author} {\bibinfo {author} {\bibfnamefont {A.}~\bibnamefont {{Aguilar-Arevalo}}} \emph {et~al.} (\bibinfo {collaboration} {DAMIC, DAMIC-M and SENSEI Collaborations}),\ }\href {https://doi.org/10.1103/PhysRevD.109.062007} {\bibfield  {journal} {\bibinfo  {journal} {Phys. Rev. D}\ }\textbf {\bibinfo {volume} {109}},\ \bibinfo {pages} {062007} (\bibinfo {year} {2024})}\BibitemShut {NoStop}%
\bibitem [{\citenamefont {Lindhard}\ \emph {et~al.}(1963)\citenamefont {Lindhard} \emph {et~al.}}]{lindhard1963integral}%
  \BibitemOpen
  \bibfield  {author} {\bibinfo {author} {\bibfnamefont {J.}~\bibnamefont {Lindhard}} \emph {et~al.},\ }\href {https://www.osti.gov/biblio/4701226} {\bibfield  {journal} {\bibinfo  {journal} {Mat. Fys. Medd. Dan. Vid. Selsk}\ }\textbf {\bibinfo {volume} {33}},\ \bibinfo {pages} {1} (\bibinfo {year} {1963})}\BibitemShut {NoStop}%
\bibitem [{\citenamefont {Tiffenberg}\ \emph {et~al.}(2017)\citenamefont {Tiffenberg} \emph {et~al.}}]{tiffenbergSingleElectronSinglePhotonSensitivity2017}%
  \BibitemOpen
  \bibfield  {author} {\bibinfo {author} {\bibfnamefont {J.}~\bibnamefont {Tiffenberg}} \emph {et~al.},\ }\href {https://doi.org/10.1103/PhysRevLett.119.131802} {\bibfield  {journal} {\bibinfo  {journal} {Phys. Rev. Lett.}\ }\textbf {\bibinfo {volume} {119}},\ \bibinfo {pages} {131802} (\bibinfo {year} {2017})}\BibitemShut {NoStop}%
\bibitem [{\citenamefont {Perez}\ \emph {et~al.}(2024)\citenamefont {Perez} \emph {et~al.}}]{perezSearchingMillichargedParticles2024}%
  \BibitemOpen
  \bibfield  {author} {\bibinfo {author} {\bibfnamefont {S.}~\bibnamefont {Perez}} \emph {et~al.} (\bibinfo {collaboration} {Oscura Collaboration}),\ }\href {https://doi.org/10.1007/JHEP02(2024)072} {\bibfield  {journal} {\bibinfo  {journal} {J. High Energ. Phys.}\ }\textbf {\bibinfo {volume} {2024}}\bibinfo  {number} { (2)},\ \bibinfo {pages} {72}}\BibitemShut {NoStop}%
\bibitem [{\citenamefont {Albakry}\ \emph {et~al.}(2023)\citenamefont {Albakry} \emph {et~al.}}]{albakryFirstMeasurementNuclearRecoil2023}%
  \BibitemOpen
\bibfield  {number} {  }\bibfield  {author} {\bibinfo {author} {\bibfnamefont {M.~F.}\ \bibnamefont {Albakry}} \emph {et~al.} (\bibinfo {collaboration} {SuperCDMS Collaboration}),\ }\href {https://doi.org/10.1103/PhysRevLett.131.091801} {\bibfield  {journal} {\bibinfo  {journal} {Phys. Rev. Lett.}\ }\textbf {\bibinfo {volume} {131}},\ \bibinfo {pages} {091801} (\bibinfo {year} {2023})}\BibitemShut {NoStop}%
\bibitem [{\citenamefont {Sorensen}(2015)}]{sorensenAtomicLimitsSearch2015}%
  \BibitemOpen
  \bibfield  {author} {\bibinfo {author} {\bibfnamefont {P.}~\bibnamefont {Sorensen}},\ }\href {https://doi.org/10.1103/PhysRevD.91.083509} {\bibfield  {journal} {\bibinfo  {journal} {Phys. Rev. D}\ }\textbf {\bibinfo {volume} {91}},\ \bibinfo {pages} {083509} (\bibinfo {year} {2015})}\BibitemShut {NoStop}%
\bibitem [{\citenamefont {Sarkis}\ \emph {et~al.}(2020)\citenamefont {Sarkis} \emph {et~al.}}]{sarkisStudyIonizationEfficiency2020a}%
  \BibitemOpen
  \bibfield  {author} {\bibinfo {author} {\bibfnamefont {Y.}~\bibnamefont {Sarkis}} \emph {et~al.},\ }\href {https://doi.org/10.1103/PhysRevD.101.102001} {\bibfield  {journal} {\bibinfo  {journal} {Phys. Rev. D}\ }\textbf {\bibinfo {volume} {101}},\ \bibinfo {pages} {102001} (\bibinfo {year} {2020})}\BibitemShut {NoStop}%
\bibitem [{\citenamefont {Chang}\ \emph {et~al.}(2025)\citenamefont {Chang} \emph {et~al.}}]{changFirstLimitsLight2025}%
  \BibitemOpen
  \bibfield  {author} {\bibinfo {author} {\bibfnamefont {C.~L.}\ \bibnamefont {Chang}} \emph {et~al.} (\bibinfo {collaboration} {TESSERACT Collaboration}),\ }\href {https://doi.org/10.48550/ARXIV.2503.03683} {\bibinfo {title} {First {{Limits}} on {{Light Dark Matter Interactions}} in a {{Low Threshold Two Channel Athermal Phonon Detector}} from the {{TESSERACT Collaboration}}}} (\bibinfo {year} {2025})\BibitemShut {NoStop}%
\bibitem [{\citenamefont {Angloher}\ \emph {et~al.}(2024)\citenamefont {Angloher} \emph {et~al.}}]{angloherFirstObservationSingle2024}%
  \BibitemOpen
  \bibfield  {author} {\bibinfo {author} {\bibfnamefont {G.}~\bibnamefont {Angloher}} \emph {et~al.} (\bibinfo {collaboration} {CRESST Collaboration}),\ }\bibfield  {journal} {\bibinfo  {journal} {Phys. Rev. D}\ }\textbf {\bibinfo {volume} {110}},\ \href {https://doi.org/10.1103/physrevd.110.083038} {10.1103/physrevd.110.083038} (\bibinfo {year} {2024})\BibitemShut {NoStop}%
\bibitem [{\citenamefont {Goupy}\ \emph {et~al.}(2023)\citenamefont {Goupy} \emph {et~al.}}]{goupyExploringCoherentElastic2023}%
  \BibitemOpen
  \bibfield  {author} {\bibinfo {author} {\bibfnamefont {C.}~\bibnamefont {Goupy}} \emph {et~al.} (\bibinfo {collaboration} {NUCLEUS Collaboration}),\ }\bibfield  {journal} {\bibinfo  {journal} {SciPost Phys. Proc.}\ }\href {https://doi.org/10.21468/scipostphysproc.12.053} {10.21468/scipostphysproc.12.053} (\bibinfo {year} {2023})\BibitemShut {NoStop}%
\bibitem [{\citenamefont {Sarkis}\ \emph {et~al.}(2023)\citenamefont {Sarkis} \emph {et~al.}}]{sarkisIonizationEfficiencyNuclear2023}%
  \BibitemOpen
  \bibfield  {author} {\bibinfo {author} {\bibfnamefont {Y.}~\bibnamefont {Sarkis}} \emph {et~al.},\ }\href {https://doi.org/10.1103/PhysRevA.107.062811} {\bibfield  {journal} {\bibinfo  {journal} {Phys. Rev. A}\ }\textbf {\bibinfo {volume} {107}},\ \bibinfo {pages} {062811} (\bibinfo {year} {2023})}\BibitemShut {NoStop}%
\bibitem [{\citenamefont {Nordlund}\ \emph {et~al.}(2016)\citenamefont {Nordlund} \emph {et~al.}}]{nordlundLargeFractionCrystal2016}%
  \BibitemOpen
  \bibfield  {author} {\bibinfo {author} {\bibfnamefont {K.}~\bibnamefont {Nordlund}} \emph {et~al.},\ }\href {https://doi.org/10.1103/PhysRevB.94.214109} {\bibfield  {journal} {\bibinfo  {journal} {Phys. Rev. B}\ }\textbf {\bibinfo {volume} {94}},\ \bibinfo {pages} {214109} (\bibinfo {year} {2016})}\BibitemShut {NoStop}%
\bibitem [{\citenamefont {Sillanp{\"a}{\"a}}\ \emph {et~al.}(2001)\citenamefont {Sillanp{\"a}{\"a}} \emph {et~al.}}]{sillanpaaElectronicStoppingCalculated2001}%
  \BibitemOpen
  \bibfield  {author} {\bibinfo {author} {\bibfnamefont {J.}~\bibnamefont {Sillanp{\"a}{\"a}}} \emph {et~al.},\ }\href {https://doi.org/10.1103/PhysRevB.63.134113} {\bibfield  {journal} {\bibinfo  {journal} {Phys. Rev. B}\ }\textbf {\bibinfo {volume} {63}},\ \bibinfo {pages} {134113} (\bibinfo {year} {2001})}\BibitemShut {NoStop}%
\bibitem [{\citenamefont {Sillanp{\"a}{\"a}}\ \emph {et~al.}(2000)\citenamefont {Sillanp{\"a}{\"a}} \emph {et~al.}}]{sillanpaaElectronicStoppingSi2000a}%
  \BibitemOpen
  \bibfield  {author} {\bibinfo {author} {\bibfnamefont {J.}~\bibnamefont {Sillanp{\"a}{\"a}}} \emph {et~al.},\ }\href {https://doi.org/10.1103/PhysRevB.62.3109} {\bibfield  {journal} {\bibinfo  {journal} {Phys. Rev. B}\ }\textbf {\bibinfo {volume} {62}},\ \bibinfo {pages} {3109} (\bibinfo {year} {2000})}\BibitemShut {NoStop}%
\bibitem [{\citenamefont {Kitagawa}\ \emph {et~al.}(1985)\citenamefont {Kitagawa} \emph {et~al.}}]{kitagawaIonirradiationExperimentFundamental1985}%
  \BibitemOpen
  \bibfield  {author} {\bibinfo {author} {\bibfnamefont {K.}~\bibnamefont {Kitagawa}} \emph {et~al.},\ }\href {https://doi.org/10.1016/0022-3115(85)90175-8} {\bibfield  {journal} {\bibinfo  {journal} {J. Nucl. Mater.}\ }\textbf {\bibinfo {volume} {133--134}},\ \bibinfo {pages} {395} (\bibinfo {year} {1985})}\BibitemShut {NoStop}%
\bibitem [{\citenamefont {Kirk}\ \emph {et~al.}(1987)\citenamefont {Kirk} \emph {et~al.}}]{kirkCollapseDefectCascades1987}%
  \BibitemOpen
  \bibfield  {author} {\bibinfo {author} {\bibfnamefont {M.}~\bibnamefont {Kirk}} \emph {et~al.},\ }\href {https://doi.org/10.1016/0022-3115(87)90494-6} {\bibfield  {journal} {\bibinfo  {journal} {J. Nucl. Mater.}\ }\textbf {\bibinfo {volume} {149}},\ \bibinfo {pages} {21} (\bibinfo {year} {1987})}\BibitemShut {NoStop}%
\bibitem [{\citenamefont {Nordlund}\ \emph {et~al.}(1999)\citenamefont {Nordlund} \emph {et~al.}}]{nordlundFormationStackingfaultTetrahedra1999}%
  \BibitemOpen
  \bibfield  {author} {\bibinfo {author} {\bibfnamefont {K.}~\bibnamefont {Nordlund}} \emph {et~al.},\ }\href {https://doi.org/10.1063/1.123948} {\bibfield  {journal} {\bibinfo  {journal} {Appl. Phys. Lett.}\ }\textbf {\bibinfo {volume} {74}},\ \bibinfo {pages} {2720} (\bibinfo {year} {1999})}\BibitemShut {NoStop}%
\bibitem [{\citenamefont {Timonova}\ \emph {et~al.}(2011)\citenamefont {Timonova} \emph {et~al.}}]{timonovaMolecularDynamicsSimulations2011}%
  \BibitemOpen
  \bibfield  {author} {\bibinfo {author} {\bibfnamefont {M.}~\bibnamefont {Timonova}} \emph {et~al.},\ }\href {https://doi.org/10.1016/j.commatsci.2011.03.016} {\bibfield  {journal} {\bibinfo  {journal} {Comput. Mater. Sci.}\ }\textbf {\bibinfo {volume} {50}},\ \bibinfo {pages} {2380} (\bibinfo {year} {2011})}\BibitemShut {NoStop}%
\bibitem [{\citenamefont {Nord}\ \emph {et~al.}(2002)\citenamefont {Nord} \emph {et~al.}}]{nordAmorphizationMechanismDefect2002}%
  \BibitemOpen
  \bibfield  {author} {\bibinfo {author} {\bibfnamefont {J.}~\bibnamefont {Nord}} \emph {et~al.},\ }\href {https://doi.org/10.1103/PhysRevB.65.165329} {\bibfield  {journal} {\bibinfo  {journal} {Phys. Rev. B}\ }\textbf {\bibinfo {volume} {65}},\ \bibinfo {pages} {165329} (\bibinfo {year} {2002})}\BibitemShut {NoStop}%
\bibitem [{\citenamefont {Thompson}\ \emph {et~al.}(2022)\citenamefont {Thompson} \emph {et~al.}}]{thompsonLAMMPSFlexibleSimulation2022}%
  \BibitemOpen
  \bibfield  {author} {\bibinfo {author} {\bibfnamefont {A.~P.}\ \bibnamefont {Thompson}} \emph {et~al.},\ }\href {https://doi.org/10.1016/j.cpc.2021.108171} {\bibfield  {journal} {\bibinfo  {journal} {Comput. Phys. Commun.}\ }\textbf {\bibinfo {volume} {271}},\ \bibinfo {pages} {108171} (\bibinfo {year} {2022})}\BibitemShut {NoStop}%
\bibitem [{\citenamefont {Brown}\ and\ \citenamefont {Yamada}(2013)}]{brownImplementingMolecularDynamics2013}%
  \BibitemOpen
  \bibfield  {author} {\bibinfo {author} {\bibfnamefont {W.~M.}\ \bibnamefont {Brown}}\ and\ \bibinfo {author} {\bibfnamefont {M.}~\bibnamefont {Yamada}},\ }\href {https://doi.org/10.1016/j.cpc.2013.08.002} {\bibfield  {journal} {\bibinfo  {journal} {Comput. Phys. Commun.}\ }\textbf {\bibinfo {volume} {184}},\ \bibinfo {pages} {2785} (\bibinfo {year} {2013})}\BibitemShut {NoStop}%
\bibitem [{\citenamefont {Devanathan}(2020)}]{devanathanInteratomicPotentialsNuclear2020}%
  \BibitemOpen
  \bibfield  {author} {\bibinfo {author} {\bibfnamefont {R.}~\bibnamefont {Devanathan}},\ }in\ \href {https://doi.org/10.1007/978-3-319-44680-6_118} {\emph {\bibinfo {booktitle} {Handbook of {{Materials Modeling}}}}},\ \bibinfo {editor} {edited by\ \bibinfo {editor} {\bibfnamefont {W.}~\bibnamefont {Andreoni}}\ and\ \bibinfo {editor} {\bibfnamefont {S.}~\bibnamefont {Yip}}}\ (\bibinfo  {publisher} {Springer International Publishing},\ \bibinfo {address} {Cham},\ \bibinfo {year} {2020})\ pp.\ \bibinfo {pages} {2141--2159}\BibitemShut {NoStop}%
\bibitem [{\citenamefont {Tersoff}(1988)}]{tersoffNewEmpiricalApproach1988}%
  \BibitemOpen
  \bibfield  {author} {\bibinfo {author} {\bibfnamefont {J.}~\bibnamefont {Tersoff}},\ }\href {https://doi.org/10.1103/PhysRevB.37.6991} {\bibfield  {journal} {\bibinfo  {journal} {Phys. Rev. B}\ }\textbf {\bibinfo {volume} {37}},\ \bibinfo {pages} {6991} (\bibinfo {year} {1988})}\BibitemShut {NoStop}%
\bibitem [{\citenamefont {Nguyen}(2017)}]{nguyenGPUacceleratedTersoffPotentials2017}%
  \BibitemOpen
  \bibfield  {author} {\bibinfo {author} {\bibfnamefont {T.~D.}\ \bibnamefont {Nguyen}},\ }\href {https://doi.org/10.1016/j.cpc.2016.10.020} {\bibfield  {journal} {\bibinfo  {journal} {Comput. Phys. Commun.}\ }\textbf {\bibinfo {volume} {212}},\ \bibinfo {pages} {113} (\bibinfo {year} {2017})}\BibitemShut {NoStop}%
\bibitem [{\citenamefont {Miller}\ \emph {et~al.}(1994)\citenamefont {Miller} \emph {et~al.}}]{millerDisplacementthresholdEnergiesSi1994}%
  \BibitemOpen
  \bibfield  {author} {\bibinfo {author} {\bibfnamefont {L.~A.}\ \bibnamefont {Miller}} \emph {et~al.},\ }\href {https://doi.org/10.1103/PhysRevB.49.16953} {\bibfield  {journal} {\bibinfo  {journal} {Phys. Rev. B}\ }\textbf {\bibinfo {volume} {49}},\ \bibinfo {pages} {16953} (\bibinfo {year} {1994})}\BibitemShut {NoStop}%
\bibitem [{\citenamefont {Farid}\ \emph {et~al.}(1991)\citenamefont {Farid} \emph {et~al.}}]{faridCohesiveEnergiesCrystals1991}%
  \BibitemOpen
  \bibfield  {author} {\bibinfo {author} {\bibfnamefont {B.}~\bibnamefont {Farid}} \emph {et~al.},\ }\href {https://doi.org/10.1103/PhysRevB.43.14248} {\bibfield  {journal} {\bibinfo  {journal} {Phys. Rev. B}\ }\textbf {\bibinfo {volume} {43}},\ \bibinfo {pages} {14248} (\bibinfo {year} {1991})}\BibitemShut {NoStop}%
\bibitem [{\citenamefont {Devanathan}\ \emph {et~al.}(1998)\citenamefont {Devanathan} \emph {et~al.}}]{devanathanDisplacementThresholdEnergies1998}%
  \BibitemOpen
  \bibfield  {author} {\bibinfo {author} {\bibfnamefont {R.}~\bibnamefont {Devanathan}} \emph {et~al.},\ }\href {https://doi.org/10.1016/S0022-3115(97)00304-8} {\bibfield  {journal} {\bibinfo  {journal} {J. Nucl. Mater.}\ }\textbf {\bibinfo {volume} {253}},\ \bibinfo {pages} {47} (\bibinfo {year} {1998})}\BibitemShut {NoStop}%
\bibitem [{\citenamefont {Ziegler}\ \emph {et~al.}(2015)\citenamefont {Ziegler} \emph {et~al.}}]{zieglerSRIMStoppingRange2015}%
  \BibitemOpen
  \bibfield  {author} {\bibinfo {author} {\bibfnamefont {J.~F.}\ \bibnamefont {Ziegler}} \emph {et~al.},\ }\href {http://www.srim.org} {\emph {\bibinfo {title} {{{SRIM}} - the Stopping and Range of Ions in Matter}}}\ (\bibinfo  {publisher} {SRIM},\ \bibinfo {address} {Chester, Maryland},\ \bibinfo {year} {2015})\BibitemShut {NoStop}%
\bibitem [{\citenamefont {Lohmann}\ \emph {et~al.}(2020)\citenamefont {Lohmann} \emph {et~al.}}]{lohmannTrajectorydependentElectronicExcitations2020}%
  \BibitemOpen
  \bibfield  {author} {\bibinfo {author} {\bibfnamefont {S.}~\bibnamefont {Lohmann}} \emph {et~al.},\ }\href {https://doi.org/10.1103/PhysRevA.102.062803} {\bibfield  {journal} {\bibinfo  {journal} {Phys. Rev. A}\ }\textbf {\bibinfo {volume} {102}},\ \bibinfo {pages} {062803} (\bibinfo {year} {2020})}\BibitemShut {NoStop}%
\bibitem [{\citenamefont {Lim}\ \emph {et~al.}(2016)\citenamefont {Lim} \emph {et~al.}}]{limElectronElevatorExcitations2016}%
  \BibitemOpen
  \bibfield  {author} {\bibinfo {author} {\bibfnamefont {A.}~\bibnamefont {Lim}} \emph {et~al.},\ }\href {https://doi.org/10.1103/PhysRevLett.116.043201} {\bibfield  {journal} {\bibinfo  {journal} {Phys. Rev. Lett.}\ }\textbf {\bibinfo {volume} {116}},\ \bibinfo {pages} {043201} (\bibinfo {year} {2016})}\BibitemShut {NoStop}%
\bibitem [{\citenamefont {Izraelevitch}\ \emph {et~al.}(2017)\citenamefont {Izraelevitch} \emph {et~al.}}]{izraelevitchMeasurementIonizationEfficiency2017a}%
  \BibitemOpen
  \bibfield  {author} {\bibinfo {author} {\bibfnamefont {F.}~\bibnamefont {Izraelevitch}} \emph {et~al.},\ }\href {https://doi.org/10.1088/1748-0221/12/06/P06014} {\bibfield  {journal} {\bibinfo  {journal} {J. Inst.}\ }\textbf {\bibinfo {volume} {12}},\ \bibinfo {pages} {P06014} (\bibinfo {year} {2017})}\BibitemShut {NoStop}%
\bibitem [{\citenamefont {Holmstr{\"o}m}\ \emph {et~al.}(2008)\citenamefont {Holmstr{\"o}m} \emph {et~al.}}]{holmstromThresholdDefectProduction2008}%
  \BibitemOpen
  \bibfield  {author} {\bibinfo {author} {\bibfnamefont {E.}~\bibnamefont {Holmstr{\"o}m}} \emph {et~al.},\ }\href {https://doi.org/10.1103/PhysRevB.78.045202} {\bibfield  {journal} {\bibinfo  {journal} {Phys. Rev. B}\ }\textbf {\bibinfo {volume} {78}},\ \bibinfo {pages} {045202} (\bibinfo {year} {2008})}\BibitemShut {NoStop}%
\bibitem [{\citenamefont {Zecher}\ \emph {et~al.}(1990)\citenamefont {Zecher} \emph {et~al.}}]{zecherEnergyDepositionEnergetic1990}%
  \BibitemOpen
  \bibfield  {author} {\bibinfo {author} {\bibfnamefont {P.}~\bibnamefont {Zecher}} \emph {et~al.},\ }\href {https://doi.org/10.1103/PhysRevA.41.4058} {\bibfield  {journal} {\bibinfo  {journal} {Phys. Rev. A}\ }\textbf {\bibinfo {volume} {41}},\ \bibinfo {pages} {4058} (\bibinfo {year} {1990})}\BibitemShut {NoStop}%
\bibitem [{\citenamefont {Gerbier}\ \emph {et~al.}(1990)\citenamefont {Gerbier} \emph {et~al.}}]{gerbierMeasurementIonizationSlow1990}%
  \BibitemOpen
  \bibfield  {author} {\bibinfo {author} {\bibfnamefont {G.}~\bibnamefont {Gerbier}} \emph {et~al.},\ }\href {https://doi.org/10.1103/PhysRevD.42.3211} {\bibfield  {journal} {\bibinfo  {journal} {Phys. Rev. D}\ }\textbf {\bibinfo {volume} {42}},\ \bibinfo {pages} {3211} (\bibinfo {year} {1990})}\BibitemShut {NoStop}%
\bibitem [{\citenamefont {Chavarria}\ \emph {et~al.}(2016)\citenamefont {Chavarria} \emph {et~al.}}]{chavarriaMeasurementIonizationProduced2016}%
  \BibitemOpen
  \bibfield  {author} {\bibinfo {author} {\bibfnamefont {A.~E.}\ \bibnamefont {Chavarria}} \emph {et~al.},\ }\href {https://doi.org/10.1103/PhysRevD.94.082007} {\bibfield  {journal} {\bibinfo  {journal} {Phys. Rev. D}\ }\textbf {\bibinfo {volume} {94}},\ \bibinfo {pages} {082007} (\bibinfo {year} {2016})}\BibitemShut {NoStop}%
\bibitem [{\citenamefont {Dougherty}(1992)}]{doughertyMeasurementsIonizationProduced1992}%
  \BibitemOpen
  \bibfield  {author} {\bibinfo {author} {\bibfnamefont {B.~L.}\ \bibnamefont {Dougherty}},\ }\href {https://doi.org/10.1103/PhysRevA.45.2104} {\bibfield  {journal} {\bibinfo  {journal} {Phys. Rev. A}\ }\textbf {\bibinfo {volume} {45}},\ \bibinfo {pages} {2104} (\bibinfo {year} {1992})}\BibitemShut {NoStop}%
\bibitem [{\citenamefont {Agnese}\ \emph {et~al.}(2018)\citenamefont {Agnese} \emph {et~al.}}]{agneseNuclearrecoilEnergyScale2018}%
  \BibitemOpen
  \bibfield  {author} {\bibinfo {author} {\bibfnamefont {R.}~\bibnamefont {Agnese}} \emph {et~al.},\ }\href {https://doi.org/10.1016/j.nima.2018.07.028} {\bibfield  {journal} {\bibinfo  {journal} {Nucl. Instrum. Meth. A}\ }\textbf {\bibinfo {volume} {905}},\ \bibinfo {pages} {71} (\bibinfo {year} {2018})}\BibitemShut {NoStop}%
\bibitem [{\citenamefont {Romani}\ \emph {et~al.}(2018)\citenamefont {Romani} \emph {et~al.}}]{romaniThermalDetectionSingle2018}%
  \BibitemOpen
  \bibfield  {author} {\bibinfo {author} {\bibfnamefont {R.~K.}\ \bibnamefont {Romani}} \emph {et~al.},\ }\href {https://doi.org/10.1063/1.5010699} {\bibfield  {journal} {\bibinfo  {journal} {Appl. Phys. Lett.}\ }\textbf {\bibinfo {volume} {112}},\ \bibinfo {pages} {043501} (\bibinfo {year} {2018})}\BibitemShut {NoStop}%
\bibitem [{\citenamefont {Fano}(1947)}]{fanoIonizationYieldRadiations1947}%
  \BibitemOpen
  \bibfield  {author} {\bibinfo {author} {\bibfnamefont {U.}~\bibnamefont {Fano}},\ }\href {https://doi.org/10.1103/PhysRev.72.26} {\bibfield  {journal} {\bibinfo  {journal} {Phys. Rev.}\ }\textbf {\bibinfo {volume} {72}},\ \bibinfo {pages} {26} (\bibinfo {year} {1947})}\BibitemShut {NoStop}%
\bibitem [{\citenamefont {Matheny}\ \emph {et~al.}(2022)\citenamefont {Matheny} \emph {et~al.}}]{mathenyIntrinsicFanoFactor2022}%
  \BibitemOpen
  \bibfield  {author} {\bibinfo {author} {\bibfnamefont {M.}~\bibnamefont {Matheny}} \emph {et~al.},\ }\href {https://doi.org/10.1103/PhysRevD.106.123009} {\bibfield  {journal} {\bibinfo  {journal} {Phys. Rev. D}\ }\textbf {\bibinfo {volume} {106}},\ \bibinfo {pages} {123009} (\bibinfo {year} {2022})}\BibitemShut {NoStop}%
\bibitem [{\citenamefont {Mazziotta}(2008)}]{mazziottaElectronHolePair2008}%
  \BibitemOpen
  \bibfield  {author} {\bibinfo {author} {\bibfnamefont {M.}~\bibnamefont {Mazziotta}},\ }\href {https://doi.org/10.1016/j.nima.2007.10.043} {\bibfield  {journal} {\bibinfo  {journal} {Nucl. Instrum. Meth. A}\ }\textbf {\bibinfo {volume} {584}},\ \bibinfo {pages} {436} (\bibinfo {year} {2008})}\BibitemShut {NoStop}%
\bibitem [{\citenamefont {Bousselham}\ \emph {et~al.}(2010)\citenamefont {Bousselham} \emph {et~al.}}]{bousselhamPhotoelectronAnticorrelationsSubPoisson2010}%
  \BibitemOpen
  \bibfield  {author} {\bibinfo {author} {\bibfnamefont {A.}~\bibnamefont {Bousselham}} \emph {et~al.},\ }\href {https://doi.org/10.1016/j.nima.2010.03.152} {\bibfield  {journal} {\bibinfo  {journal} {Nucl. Instrum. Meth. A}\ }\textbf {\bibinfo {volume} {620}},\ \bibinfo {pages} {359} (\bibinfo {year} {2010})}\BibitemShut {NoStop}%
\bibitem [{\citenamefont {Baxter}\ \emph {et~al.}(2021)\citenamefont {Baxter} \emph {et~al.}}]{baxterRecommendedConventionsReporting2021}%
  \BibitemOpen
  \bibfield  {author} {\bibinfo {author} {\bibfnamefont {D.}~\bibnamefont {Baxter}} \emph {et~al.},\ }\href {https://doi.org/10.1140/epjc/s10052-021-09655-y} {\bibfield  {journal} {\bibinfo  {journal} {Eur. Phys. J. C}\ }\textbf {\bibinfo {volume} {81}},\ \bibinfo {pages} {907} (\bibinfo {year} {2021})}\BibitemShut {NoStop}%
\bibitem [{\citenamefont {Savage}\ \emph {et~al.}(2009)\citenamefont {Savage} \emph {et~al.}}]{savageCompatibilityDAMALIBRA2009}%
  \BibitemOpen
  \bibfield  {author} {\bibinfo {author} {\bibfnamefont {C.}~\bibnamefont {Savage}} \emph {et~al.},\ }\href {https://doi.org/10.1088/1475-7516/2009/04/010} {\bibfield  {journal} {\bibinfo  {journal} {J. Cosmol. Astropart. Phys.}\ }\textbf {\bibinfo {volume} {2009}}\bibinfo  {number} { (04)},\ \bibinfo {pages} {010}}\BibitemShut {NoStop}%
\bibitem [{\citenamefont {Adari}\ \emph {et~al.}(2025)\citenamefont {Adari} \emph {et~al.}}]{adariFirstDirectDetectionResults2025}%
  \BibitemOpen
\bibfield  {number} {  }\bibfield  {author} {\bibinfo {author} {\bibfnamefont {P.}~\bibnamefont {Adari}} \emph {et~al.} (\bibinfo {collaboration} {SENSEI Collaboration}),\ }\href {https://doi.org/10.1103/PhysRevLett.134.011804} {\bibfield  {journal} {\bibinfo  {journal} {Phys. Rev. Lett.}\ }\textbf {\bibinfo {volume} {134}},\ \bibinfo {pages} {011804} (\bibinfo {year} {2025})}\BibitemShut {NoStop}%
\bibitem [{\citenamefont {Albakry}\ \emph {et~al.}(2022)\citenamefont {Albakry} \emph {et~al.}}]{albakryInvestigatingSourcesLowenergy2022}%
  \BibitemOpen
  \bibfield  {author} {\bibinfo {author} {\bibfnamefont {M.~F.}\ \bibnamefont {Albakry}} \emph {et~al.} (\bibinfo {collaboration} {SuperCDMS Collaboration}),\ }\href {https://doi.org/10.1103/PhysRevD.105.112006} {\bibfield  {journal} {\bibinfo  {journal} {Phys. Rev. D}\ }\textbf {\bibinfo {volume} {105}},\ \bibinfo {pages} {112006} (\bibinfo {year} {2022})}\BibitemShut {NoStop}%
\bibitem [{\citenamefont {Alkhatib}\ \emph {et~al.}(2021)\citenamefont {Alkhatib} \emph {et~al.}}]{alkhatibLightDarkMatter2021}%
  \BibitemOpen
  \bibfield  {author} {\bibinfo {author} {\bibfnamefont {I.}~\bibnamefont {Alkhatib}} \emph {et~al.} (\bibinfo {collaboration} {SuperCDMS Collaboration}),\ }\href {https://doi.org/10.1103/PhysRevLett.127.061801} {\bibfield  {journal} {\bibinfo  {journal} {Phys. Rev. Lett.}\ }\textbf {\bibinfo {volume} {127}},\ \bibinfo {pages} {061801} (\bibinfo {year} {2021})}\BibitemShut {NoStop}%
\bibitem [{\citenamefont {Angloher}\ \emph {et~al.}(2023)\citenamefont {Angloher} \emph {et~al.}}]{angloherResultsSubGeVDark2023}%
  \BibitemOpen
  \bibfield  {author} {\bibinfo {author} {\bibfnamefont {G.}~\bibnamefont {Angloher}} \emph {et~al.} (\bibinfo {collaboration} {CRESST Collaboration}),\ }\href {https://doi.org/10.1103/PhysRevD.107.122003} {\bibfield  {journal} {\bibinfo  {journal} {Phys. Rev. D}\ }\textbf {\bibinfo {volume} {107}},\ \bibinfo {pages} {122003} (\bibinfo {year} {2023})}\BibitemShut {NoStop}%
\bibitem [{\citenamefont {{Aguilar-Arevalo}}\ \emph {et~al.}(2020)\citenamefont {{Aguilar-Arevalo}} \emph {et~al.}}]{aguilar-arevaloResultsLowMassWeakly2020}%
  \BibitemOpen
  \bibfield  {author} {\bibinfo {author} {\bibfnamefont {A.}~\bibnamefont {{Aguilar-Arevalo}}} \emph {et~al.} (\bibinfo {collaboration} {DAMIC Collaboration}),\ }\href {https://doi.org/10.1103/PhysRevLett.125.241803} {\bibfield  {journal} {\bibinfo  {journal} {Phys. Rev. Lett.}\ }\textbf {\bibinfo {volume} {125}},\ \bibinfo {pages} {241803} (\bibinfo {year} {2020})}\BibitemShut {NoStop}%
\end{thebibliography}%
\end{document}